\documentclass[aps,pra,twocolumn,groupedaddress,showpacs]{revtex4}

\begin{document}


\title{Rotational structures of long-range diatomic molecules}


\author{Bo Gao}
\email[]{bgao@physics.utoledo.edu}
\homepage[]{http://bgaowww.physics.utoledo.edu}
\affiliation{Department of Physics and Astronomy,
	University of Toledo,
	Toledo, Ohio 43606}


\date{June 15, 2004}

\begin{abstract}

We present a systematic understanding of the rotational structure
of a long-range (vibrationally highly-excited) diatomic molecule.
For example, we show that depending on a quantum defect, 
the least-bound vibrational state of a diatomic molecule with 
$-C_n/r^n$ ($n>2$) asymptotic interaction can have only 1, 2, and 
up to a maximum of $n-2$ rotational states. A classification scheme 
of diatomic molecules is proposed, in which each class has a distinctive 
rotational structure and corresponds to different atom-atom 
scattering properties above the dissociation limit.

\end{abstract}


\maketitle


\section{Introduction}

How fast can we rotate a molecule before breaking it \cite{kar99,li00}? 
How does a rotational series terminates at the dissociation limit?
How many rotational levels are there for a diatomic molecule in its
last (least-bound), or next-to-last, vibrational state?
These intriguing, and closely related, questions are taking on 
a new dimension of practical importance as our ability to make 
large samples of long-range molecules (vibrational highly-excited molecules) 
\cite{stw78,tho87,stw76,tie93,don02,reg03,xu03,her03,cub03,str03,joc03a}, 
and even condensates of long-range molecules \cite{gre03,joc03b,zwi03}, 
continues to grow.
To understand the properties of a long-range molecule,
especially how it responds to external perturbations such as
collision with other atoms,
we need not only the properties of a particular molecular
state, such as the least-bound $s$ state. We also need to know 
what are the states around it. It is this global structure of states
that is the focus of this work.

One approach to this problem is to compute the universal
spectra for each type of long-range interaction, $-C_n/r^n$, as we have done
previously for $n=6$ \cite{gao01,gao00} and $n=3$ \cite{gao99b}, and simply
observe them. This would, however, be very tedious and can never 
be completely inclusive. Our approach here is based on the
recognition that the global structure of states, not including
specific values for binding energies, depends only on the
zero-energy wave function, more specifically, on
its number of nodes as a function of both the angular momentum 
quantum number $l$ and the exponent $n$ characterizing the long-range
interaction.

Our results, and answers to the questions raised above, can be 
summarized in two simple formulas that will be derived later in
the article. The first gives the dependence of the number of bound
states on the angular momentum $l$ for a quantum system with 
$-C_n/r^n$ ($n>2$) long-range interaction.
\begin{equation}
N_l = \left[N_0+\mu^c-\frac{1}{n-2}l\right] \;.
\label{eq:Nlu}
\end{equation}
Here $[x]$ means the greatest integer less than or equal to $x$.
$N_l$ is the number of bound states of angular momentum $l$.
$N_0$ is the number of $s$ wave bound states.
$\mu^c$ is a quantum defect, to be defined later, that has a range
of $0\le \mu^c<1$.

The second formula relates the quantum defect to the $s$ wave scattering
length
\begin{equation}
a_{0s} = \bar{a}_{0s}
	\frac{\tan(\pi\mu^c) + \tan(\pi b)}{\tan(\pi\mu^c)} \;.
\label{eq:a0smuc1}
\end{equation}
Here $a_{0s}=a_0/\beta_n$ is the $s$ wave scattering length, $a_0$, scaled by the length
scale $\beta_n = (2\mu C_n/\hbar^2)^{1/(n-2)}$ associated with the long-range interaction; 
$b=1/(n-2)$; and
\begin{equation}
\bar{a}_{0s} = \cos(\pi b)\left[b^{2b}\frac{\Gamma(1-b)}{\Gamma(1+b)}\right] \;,
\label{eq:a0bars}
\end{equation}
is the mean $s$ wave scattering length of Gribakin and Flambaum
\cite{gri93}, scaled by $\beta_n$.

The consequences of these results are easily understood and are discussed
in Sec.~\ref{sec:class}. Equation~(\ref{eq:Nlu}) is derived in Sec.~\ref{sec:der}.
It is another example of universal properties 
at length scale $\beta_n$, as discussed in a more general
terms in two recent publications \cite{gao03a,gao04a}. 
This universal property is followed by all molecules in varying 
degrees. Deviations from it and other issues are 
discussed in Sec.~\ref{sec:dis}. A primer of 
the angular-momentum-insensitive quantum-defect theory 
(AQDT) \cite{gao01,gao00}, which is the foundation of 
this work, can be found in Appendix~\ref{sec:aqdtp}.

\section{Derivation of Equation~(\ref{eq:Nlu})}
\label{sec:der}

Equation~(\ref{eq:Nlu}) may be derived using two different methods. One is
to apply AQDT \cite{gao01,gao00}, the version for arbitrary $n>2$ 
as outline in \cite{gao03a} and Appendix~\ref{sec:aqdtp},
to the zero-energy state of a diatomic molecule. This approach is
discussed briefly in Appendix~\ref{sec:aqdtp}. The other approach
is the method of effective potential \cite{gao03a,gao04a}.
It is this latter method that we present here, for the purpose
of further promoting this powerful concept.
While it makes no difference in this particular case, 
for more complex systems, such as quantum few-body
or quantum many-body systems \cite{gao04a} where no analytic solutions are
available, the method of effective potential may be the only way to uncover 
universal properties at different length scales. The results would, 
of course, be mostly numerical in those cases. But a numerical 
solution done right can indeed yield universal behavior \cite{gao04a}.

The method of effective potential is very simple.
It states that for a physical observable that depends only on 
states around the threshold, such as the number of nodes of the zero-energy 
wave function that we are looking at here, its universal behavior at 
length scale $\beta_n$ can be derived from
any potential that has the right asymptotic behavior and 
is strongly repulsive at short distances. Specifically, 
a universal result at length scale $\beta_n$ is obtained from the
corresponding result for the effective potential by taking a
proper limit to eliminate the shorter length scales 
while keeping the short-range K matrix, 
$K^c(0,l)$ (\cite{gao01} and Appendix~\ref{sec:aqdtp}), to be
a constant for one particular $l$ \cite{gao03a,gao04a}.
	
We take here, for simplicity, a hard-sphere with
an attractive tail (HST), 
\begin{equation}
V_{\text{HST}}(r) = \left\{ \begin{array}{lcl}
	\infty & , & r \le r_0 \\
	-C_n/r^n & , & r>r_0 \end{array} \right. \;,
\label{eq:HST}
\end{equation}
as our effective potential.
Its number of bound levels for angular momentum $l$
is given by (\cite{gao03a} and Appendix~\ref{sec:aqdtp})
\begin{equation}
N_{\text{HST}}(l) = \left\{ \begin{array}{lcl}
	m & , & j_{\nu_0,m} \le y_0 < j_{\nu_0,m+1} \\
	0 & , & y_0 < j_{\nu_0,1}
	\end{array} \right.\;,
\label{eq:NlHST}	
\end{equation}
where $y_0=[2/(n-2)](\beta_n/r_0)^{(n-2)/2}$,
$\nu_0=(2l+1)/(n-2)$, and
$j_{\nu_0,m}$ ($m\ge 1$) is the m-th zero of the 
Bessel function $J_{\nu_0}(x)$ \cite{abr64}.

Its $K^c$ parameter at zero energy is 
given by (\cite{gao03a} and Appendix~\ref{sec:aqdtp}).
\begin{equation}
K_{\text{HST}}^c(0,l)
	= -\frac{J_{\nu_0}(y_0)\cos(\pi\nu_0/2)-Y_{\nu_0}(y_0)\sin(\pi\nu_0/2)}
		{J_{\nu_0}(y_0)\sin(\pi\nu_0/2)+Y_{\nu_0}(y_0)\cos(\pi\nu_0/2)} ,
\label{eq:KcHST}		
\end{equation}
where $J$ and $Y$ are the Bessel functions \cite{abr64}.

In the limit of $r_0\rightarrow 0+$ that eliminates the shorter length scale
(see \cite{gao04a} for a more precise 
definition), $y_0\gg 1$, and the roots of the Bessel function are 
given by \cite{wat96}
\begin{equation}
j_{\nu_0,m} \rightarrow (m+\nu_0/2-1/4)\pi \;.
\end{equation}
$K^c(0,l)$ becomes an $l$-independent constant
\begin{equation}
K^c(0,l)\rightarrow K^c = \tan(y_0+\pi/4) \;.
\end{equation}
Define the quantum defect, $\mu^c(\epsilon,l)$, to be a parameter in a
range of $0\le \mu^c<1$ and related to $K^c$ by
\begin{equation}
K^c(\epsilon,l) = \tan\left[\pi\mu^c(\epsilon,l)+\frac{\pi}{2(n-2)}\right] \;.
\label{eq:mucdef}
\end{equation}
It is clear that $\mu^c(0,l)$ also becomes an $l$-independent constant
\begin{equation}
\mu^c(0,l)\rightarrow \mu^c = \frac{y_0}
	{\pi}+\frac{1}{4}-\frac{1}{2(n-2)}-j \;,
\end{equation}
where $j$ is an integer chosen such that $\mu^c$ falls in the range of
$0\le \mu^c<1$. 

Combining these results, the number of bound states of 
angular momentum $l$ can be written in the limit of 
$r_0\rightarrow 0+$ as
\begin{equation}
N_{\text{HST}}(l) \rightarrow \begin{array}{lcl}
	m & , & m \le j+\mu^c-\frac{1}{n-2}l < m+1 \end{array} \;.
\end{equation}
In other words,
\begin{equation}
N_{\text{HST}}(l) \stackrel{r_0\rightarrow 0+}
	{\longrightarrow}\left[j+\mu^c-\frac{1}{n-2}l\right] \;.
\label{eq:NlHSTL}	
\end{equation}
where $[x]$ again means the greatest integer less than or equal to $x$.
Note that the result on the right-hand-side of this equation is now no
longer just a property of the HST potential, but a universal
property at length scale $\beta_n$, applicable to any quantum
system with the same long-range behavior and has a $\beta_n$ that is
much longer than other length scales in the system.

Since $0\le \mu^c<1$, the integer $j$ in Eq.~(\ref{eq:NlHSTL}) 
is simply the number of
bound states for $l=0$. Eq.~(\ref{eq:Nlu}) is thus derived.
It is trivial to show that starting from a LJ(n,2n-2) effective
potential \cite{gao03a} (see also Sec.~\ref{sec:LJn})
leads to identical result. And again, the same result can
be derived by applying AQDT to the zero-energy diatomic 
state, as outlined in Appendix~\ref{sec:aqdtp}.

We emphasize that the method of effective potential is fundamentally
different from the pseudopotential method \cite{hua57}.
In the latter method, a different pseudopotential is required for
each partial wave. And each pseudopotential has at least one 
independent parameter to characterize the scattering of that 
particular partial wave.
Without another theory relating scattering of different partial waves, 
no universal $l$-dependence of any kind can be established.
(This is in addition to its well known limitations in describing
bound states.) 
In contrast, a single effective potential is used to describe 
all $l$. And one can do this because
a single parameter in AQDT describes a multitude of angular
momentum states (see \cite{gao01,gao00} and Appendix~\ref{sec:aqdtp}).
Put it in another way. Scattering of different angular momenta
are indeed related, and so are the bound spectra of different
angular momenta. Their relationships are determined by the
long-range interaction (\cite{gao01,gao00} and Appendix~\ref{sec:aqdtp}),
and are incorporated automatically in an effective potential \cite{gao03a}. 
The universal property described by Eq.~(\ref{eq:Nlu}) is but one
reflection of this type of systematics.
To further re-enforce this concept, we leave as an exercise to derive
the relationship between the $p$ wave scattering length and
the $s$ wave scattering length for $n=6$. The derivation is trivial and
all the information required can be found in 
Appendix~\ref{sec:aqdtp} and \cite{gao98b}.
The answer will be provided in an upcoming publication.

\section{Classification of molecules using quantum defect}
\label{sec:class}

Let us first state that we do not include explicitly the effect of
statistics when atoms are identical. It would make our statements 
unnecessarily complex without introducing any new physics. 
In specific applications, all one needs to do is to exclude 
states that cannot satisfy the symmetry requirement
(see, for example, \cite{gao96}), as needed.

The physical implications of Eq.~(\ref{eq:Nlu}) can be
easily understood by noting that $N_0-1$ is \textit{the} maximum
vibrational quantum number, $v_{max}$, while $N_l-1$ is the maximum
vibrational quantum number, $v_{max,l}$, that can support 
a rotational state of angular momentum $l$.
A vibrational state $v$ can have all $l$ for which $v_{max,l}\ge v$.
Letting $L_{max,v}$ to be the maximum rotational quantum number for
vibrational state $v$, we have from Eq.~(\ref{eq:Nlu})
\begin{equation}
v = \left[v_{max}+\mu^c-\frac{1}{n-2}L_{max,v}\right] \;,
\end{equation}
which can be written as
\begin{equation}
L_{max,v} = (n-2)(v_{max}-v)+\left[(n-2)\mu^c\right] \;,
\end{equation}
where the square braket in the second term, $[x]$, again denotes 
the greatest integer less than or equal to $x$

This result suggests the classification of molecules
into $n-2$ classes, each corresponding to an equal interval of
$b=1/(n-2)$ in the quantum-defect space. 
For class $i$ with $ib\le\mu^c<(i+1)b$, 
we have $i\le (n-2)\mu^c<i+1$, and therefore
\begin{equation}
L_{max,v}(i) = i+(n-2)(v_{max}-v) \;.
\end{equation}
Thus each class of molecules corresponds to a unique rotational 
structure that terminates at $L_{max,v}(i)$.
This classification is summarized in Table~\ref{tb:class}.
In particular, it means that the least-bound vibrational state
can have 1 (Class 0), 2 (Class 1), and up to a maximum of $n-2$ (Class $n-3$)
rotational states, depending on the quantum defect of the molecule.
For the next-to-last vibrational state, add $n-2$ rotational states to
each class, and so on for lower vibrational states.

\begin{table}
\caption{Classification of diatomic molecules with $-C_n/r^n$ ($n>2$) 
long range interaction using quantum defect. Here $L_{max,v}$ is the 
maximum rotational quantum number for the vibrational state $v$. 
$\Delta v = v_{max}-v$. $a_{0s}=a_0/\beta_n$ is
the scaled $s$ wave scattering length. Note that some of the 
rotational states may be excluded for identical particles. Also note
that scattering length has no definition for $n=3$.
\label{tb:class}}
\begin{ruledtabular}
\begin{tabular}{c|c|c|c}
Class & Range of $\mu^c$ & $L_{max,v}(i)$ & Range of $a_{0s}$ \\
\hline
0 & $0 \le \mu^c < b$ & $0+(n-2)\Delta v$ & $2\bar{a}_{0s}< a_{0s}\le\infty$ \\
1 & $b \le \mu^c < 2b$ & $1+(n-2)\Delta v$ & \vdots \\
\vdots & \vdots & \vdots &\vdots \\
$n-3$ & $(n-3)b \le \mu^c < 1$ & $n-3+(n-2)\Delta v$ & $-\infty<a_{0s}\le 0$
\end{tabular}
\end{ruledtabular}
\end{table}

What makes this classification useful is that each class not only
has a distinctive rotational structure, it also corresponds to distinctive
atom-atom scattering properties above the dissociation limit.
First, each class of molecules corresponds to a distinctive 
(non-overlapping) range of scattering length, which can be determined
from Eq.~(\ref{eq:a0smuc1}) and is summarized in Table~\ref{tb:class}. 

Equation~(\ref{eq:a0smuc1}) derives easily from 
the definition of the mean scattering length \cite{gri93}, 
Eq.~(\ref{eq:a0bars}), the definition of the quantum defect, 
Eq.~(\ref{eq:mucdef}), and the following rigorous
relation between $K^c$ and the $s$ wave scattering length 
(\cite{gao03a} and Appendix~\ref{sec:aqdtp})
\begin{equation}
a_{0s} = \left[b^{2b}\frac{\Gamma(1-b)}{\Gamma(1+b)}\right]
	\frac{K^c(0,0) + \tan(\pi b/2)}{K^c(0,0) - \tan(\pi b/2)} \;,
\label{eq:a0sKc}
\end{equation}
which is similar to the relation between scattering length and
a semiclassical phase as derived by Gribakin and Flambaum
\cite{gri93}. These equations combine to give
\begin{equation}
a_{0s} = \bar{a}_{0s}
	\frac{\tan[\pi\mu^c(0,0)] + \tan(\pi b)}{\tan[\pi\mu^c(0,0)]} \;,
\label{eq:a0smuc}
\end{equation}
which is the exact relation between scattering length and quantum
defect. It is the more rigorous way to write Eq.~(\ref{eq:a0smuc1}),
applicable even when the system deviates from the universal behavior
(see Sec.~\ref{sec:dev} and Appendix~\ref{sec:aqdtp}).

With the correspondence between quantum defect and scattering length, 
our classification of molecules can translate into other general 
statements, such as, 
a) The least-bound vibrational state of a diatomic molecule with 
$a_{0s}\ge 2\bar{a}_{0s}$ has only a single rotational state.
b) The least-bound vibrational state of a diatomic molecule with negative 
scattering length has $n-2$ rotational states.
It is worth noting that molecules with negative scattering length all
fall into a single class, Class $n-3$, while molecules with positive
scattering length separate into $n-3$ classes, from Class 0 to Class
$n-4$. A similar feature was first noted by Gribakin and Flambaum
\cite{gri93}.

The different scattering properties for different classes are
not restricted to the $s$ wave. In fact, more interesting differences occur
for higher partial waves.
For example, Class 0 does not have a $p$ wave bound state for the last vibrational
level. This $p$ state, which would have been bound for $\mu^c\ge b$, does not disappear
completely. It shows itself as $p$ wave shape-resonance above the threshold,
which actually becomes infinitely narrow (infinitely long-lived) 
as one approaches $\mu^c=b$ from the
side of Class 0. In general, a Class $i$ system is the one that has a shape-resonance
of $l=i+1$ closest to the threshold. The detailed properties of these
resonances are however beyond the scope of this article 
(see, e.g. \cite{boe96,gao98b,cot99,gao01}).

The critical values of $\mu^c = ib$ that are the boundaries between different
classes correspond to having bound or quasibound states
of angular momenta $l = i+(n-2)j$ ($j$ being a non-negative integer) right at
the threshold (Appendix~\ref{sec:aqdtp}). 
They have vibrational quantum numbers of $v=v_{max}-j$, 
respectively. This is a generalization of some of the results in \cite{gao00}
to the case of arbitrary $n>3$. Note that the wave functions for zero-energy
bound or quasibound states are well defined and are given in the region of
long-range potential by (Appendix~\ref{sec:aqdtp})
\begin{equation}
u_{\epsilon=0 l}(r) = A r_s^{1/2}J_{\nu_0}(y) \;,
\label{eq:zebwfn}
\end{equation}
where $r_s=r/\beta_n$ is a scaled radius, and $y=[2/(n-2)]r_s^{-(n-2)/2}$.
This wave function has an asymptotic behavior of $1/r^l$ at large $r$,
thus representing a true, normalizable, bound state for
$l>0$, and a quasibound (not normalizable) state for
$l=0$. The fact that $s$-wave wave function in the effective-range
theory becomes completely meaningless when $a_0=\infty$ 
is only a limitation of the theory, not a reflection of any
physical reality.

\section{Discussions}
\label{sec:dis}

We discuss here some special cases, deviations from the universal
behavior, and how they might be treated.

\subsection{The case of $n=3$}

Our result is applicable to $n=3$, even though the scattering
length has no definition in this case (for any $l$) \cite{lev63,gao99a}. 
Specifically, it predicts that quantum systems with $n=3$ have 
only a single class (Class 0) with $L_{max,v} = v_{max}-v$. 
In other words, the last vibrational state for $n=3$ has a 
single rotational state, an $s$ state. The next-to-last 
vibrational state has two rotational states, and so forth.
This prediction is confirmed by the analytic solution for $-C_3/r^3$
potential \cite{gao99b,gao99a}.

\subsection{The special case of LJ(n,2n-2) potentials}
\label{sec:LJn}

For a set of Lennard-Jones potentials LJ(n,2n-2) ($n>2$) defined by
\begin{equation}
V_{\text{LJn}}(r) = -C_n/r^n+C_{2n-2}/r^{2n-2} \;,
\label{eq:LJn}
\end{equation}
the number of bound levels for any $l$ is given by 
(\cite{gao03a} and Appendix~\ref{sec:LJnDer})
\begin{equation}
N_{\text{LJn}}(l) = \left\{ \begin{array}{lcl}
	\left[z_0+\frac{1}{2}-\frac{\nu_0}{2}\right] & , & z_0\ge (\nu_0+1)/2 \\
	0 & , &  z_0<(\nu_0+1)/2 
	\end{array} \right. ,
\label{eq:NlLJn}	
\end{equation}
where $\nu_0=(2l+1)/(n-2)$, and the square braket $[x]$ again means the greatest 
integer less than or equal to $x$.
$z_0=(\beta_n/\beta_{2n-2})^{n-2}/[2(n-2)]$, where $\beta_{2n-2}$
is the length scale assoicated with the $C_{2n-2}/r^{2n-2}$ interaction.
 Thus for the LJ(n,2n-2) potential,
the universal dependence of the number of bound states on $l$,
as specified by Eq.~(\ref{eq:Nlu}),
is exact, true even when $\beta_{2n-2}$ is comparable to $\beta_n$ and 
the corresponding potentials are so shallow as to
support only a single or a few bound states.

This result implies that to break the universal dependence on $l$, one
needs not only a short-range interaction, but the behavior of this interaction
also has to be different from LJ(n,2n-2).

\subsection{Deviations from the universal behavior}
\label{sec:dev}

The key to understand qualitatively the deviations from the universal
behavior is to recognize the origin of this universality. 
The universal $l$-dependence originates from the 
$l$-independence of $K^c(0,l)$ (\cite{gao01} and Appendix~\ref{sec:aqdtp}),
which is a result of both the small mass
ratio $m_e/\mu$, where $m_e$ is the electron mass and $\mu$ is the
reduced mass of the molecule (not to be confused with the quantum
defect $\mu^c$), and the condition
of $\beta_n\gg r_0$ where $r_0$ is a representative of other length
scales in the system. (For HST potential, it coincides with the $r_0$
that we used earlier.)

With this understanding, it is clear that the universal behavior 
of Eq.~(\ref{eq:Nlu}) should be followed by all molecules
to some degree. The mass ratio $m_e/\mu$ is always small and can 
be taken for granted. (This is why we don't always mention it.) 
And almost by definition, $\beta_n$
\textit{is} the longest length scale in the problem, otherwise it would not,
and should not have been called the long-range interaction.

It is also clear that the universal behavior is best followed by the states
with highest vibrational quantum numbers. For example, consider our prediction
of $L_{max,v}(i) = i+(n-2)(v_{max}-v)$. For the least-bound vibrational
state, $v=v_{max}$, it would only require $l$-independence of $K^c$ 
over a range of $\Delta l = n-2$. In comparison, 
the same result applied to $v=v_{max}-9$
would require $l$-independence of $K^c$ over 10 times that range, 
which generally becomes considerably worse (depending also on $n$, and
other details of the short-range interaction).

As far as predictions for the last few vibrational states 
(long-range molecules) are concerned, there is no need to worry 
about deviation except when $\mu^c$ is very close to
one of the critical values of $\mu^c = ib$, where a small $l$-dependence
may mean the difference between a bound state and a shape resonance.

Whenever necessary, the deviation from universal behavior can be easily handled
within the AQDT framework. All we need is to count the nodes of the zero-energy
wave functions more carefully! As discussed in Appendix~\ref{sec:aqdtp},
AQDT is an exact formulation and an excellent platform for exact
numerical calculations. This also applies to node-counting:
integrate the Schr\"{o}dinger equation at zero energy and count the nodes up 
to a distance where $K^c(0,l)$ has converged to a desired accuracy 
[One computes $K^c(0,l)$ by matching the integrated wave function to 
that given by Eq.~(\ref{eq:wfnz1}) at different radii $r$. 
As a function this matching radius, $K^c(0,l)$ converges to a 
$r$-independent constant when the potential becomes $-C_n/r^n$ and 
the wave function becomes that of Eq.~(\ref{eq:wfnz1})]. 
Adding to that the number of nodes beyond 
this distance, which can now be calculated analytically, gives one the 
total number of nodes. This way, one would never miss a node which 
could potentially be at infinity.

One could also try to find if there are any systematics in the deviation 
by going to the next, shorter, length scale. Any such attempt would however
be necessarily system-specific and will be deferred to specific applications.
Examples of the universal rotational structure for $n=6$ can already 
be found in \cite{gao01,gao00}, though they were not discussed explicitly. 
It was the simple structures observed there that motivated this work. 

\section{Conclusion}

In conclusion, we have shown that the rotational structure of a long-range
molecule follows a simple universal behavior that is characterized by
two parameters, the exponent $n$ of the long-range interaction $-C_n/r^n$,
and a quantum defect, which is related in a simple way to the $s$ wave 
scattering length [whenever it is well defined ($n>3$)].
The resulting classification scheme gives a simple qualitative description
of both the rotational structure of a long-range molecule and the corresponding
atom-atom scattering properties above the dissociation threshold.

Finally, getting back to one of the questions at the beginning that we
have not answered explicitly: how fast can we rotate a molecule before 
breaking it? The answer is, of course, $L_{max,v}$ units of angular momenta, 
which is generally a very small number for long-range molecules.

\begin{acknowledgments}
I thank Michael Cavagnero, Eite Tesinga, Paul Julienne, and
Carl Williams for helpful discussions. 
This work was supported by the National Science Foundation under 
the Grant number PHY-0140295.
\end{acknowledgments}

\appendix

\section{AQDT: A primer}
\label{sec:aqdtp}

We give here a brief review of the angular-momentum-insensitive
quantum defect theory (AQDT) \cite{gao01,gao00}. The focus will 
be on the conceptual aspects, and issues directly related to 
this particular work. We point out that there are a number of
different quantum-defect formulations for diatomic 
systems \cite{mie80,jul89,gao96,bur98,mie00}.
There are also quantum-defect analysis \cite{kok00,kok02},
and numerical methods that incorporate
the concepts of quantum-defect theory \cite{van02}.
Only our formulation is briefly reviewed here.

Consider a radial Schr\"{o}dinger equation
\begin{equation}
\left[-\frac{\hbar^2}{2\mu}\frac{d^2}{dr^2} 
	+\frac{\hbar^2 l(l+1)}{2\mu r^2}+ V(r)-\epsilon \right]
	u_{\epsilon l}(r) = 0 \;,
\label{eq:rsch}
\end{equation}
where $V(r)$ becomes $-C_n/r^n$ beyond a distance $r_0$.
($r_0$ does not have to have a precise value like the case
of HST. It is a representative of length scales associated
with interactions of shorter range, introduced to simplify
our discussion.)

In AQDT, the wave function in the region of long-range interaction
($r\ge r_0$) is written as a linear superposition of a 
pair of reference functions
\begin{equation}
u_{\epsilon l}(r) = A_{\epsilon l}[f^c_{\epsilon_s l}(r_s) 
	- K^c(\epsilon,l) g^c_{\epsilon_s l}(r_s)]\;,
\label{eq:wfn}
\end{equation}
which also serves to define the short-range K matrix
$K^c(\epsilon,l)$. The functions $f^c$ and $g^c$ are solutions 
for the long-range potential $-C_n/r^n$ \cite{gao98a,gao99a}.
Their notations reflect the fact that with proper scaling 
and normalization, $f^c$ and $g^c$ depend on $r$ only through 
a scaled radius $r_s=r/\beta_n$ and on
energy only through a scaled energy \cite{gao98a,gao99a}
\begin{equation} 
\epsilon_s = \frac{\epsilon}{(\hbar^2/2\mu)(1/\beta_n)^2} \;.
\end{equation}
Note that for the purpose of cleaner notion for arbitrary $n$, 
we have abandoned the factor of 16 used previously for 
$n=6$ \cite{gao98a,gao98b,gao00,gao01}, and the
factor of 4 used previously for $n=3$ \cite{gao99a,gao99b}.

Much of the art of a quantum defect theory \cite{gre82,sea83,fan86,aym96}
is in choosing $f^c$ and $g^c$ that best reflect
the underlying physics. For a molecule, the wave function at
short distances is nearly independent of $l$ because the rotational
energy is small compared to electronic energy (originated from
the small mass ratio $m_e/\mu$). AQDT takes advantage of
this fact by picking a pair of solutions for $-C_n/r^n$ potential 
that have not only energy-independent, but also $l$-independent 
behavior at short distances 
(possible because $n>2$) \cite{gao01,gao03a,gao04a}:
\begin{eqnarray}
f^c_{\epsilon_s l}(r_s) &\stackrel{r\ll\beta_n}{\longrightarrow}& 
	(2/\pi)^{1/2}r_s^{n/4}
	\cos\left(y-\pi/4 \right) \;, 
\label{eq:fcasy0}\\
g^c_{\epsilon_s l}(r_s) &\stackrel{r\ll\beta_n}{\longrightarrow}& 
	-(2/\pi)^{1/2}r_s^{n/4}
	\sin\left(y -\pi/4 \right) \;,
\label{eq:gcasy0}
\end{eqnarray}
where $y=[2/(n-2)]r_s^{-(n-2)/2}$. 

With this choice of reference pairs, matching to wave function
at short distances yields an $K^c$ that is nearly independent
of $l$, provided that $r_0$ is much smaller than $\beta_n$ so
that the reference functions at this point are well represented 
by their $l$-independent form of Eqs.~(\ref{eq:fcasy0})-(\ref{eq:gcasy0}). 

An approximately $l$-independent $K^c$ thus reflects the underlying 
physics that for long-range molecules and atom-atom scattering at low 
energies, the angular momentum 
dependence is most important only at large distances where its effects   
can be incorporated analytically. This is the physical origin
of why a single parameter in AQDT is often capable of describing a multitude
of angular momentum states \cite{gao01,gao00,gao03a}. The results of this
work is but one reflection of the resulting systematics.

The approximate energy-independence of $K^c$, under the same condition
of $\beta_n\gg r_0$ is fairly standard \cite{gre82,fan86}.
It is both because the reference functions have been chosen to be 
energy-independent at short distances, and
because the short-range wave function varies with
energy on a scale of $(\hbar^2/2\mu)(1/r_0)^2$, much greater
than the corresponding energy scale associated with the long-range
interaction, which is $(\hbar^2/2\mu)(1/\beta_n)^2$ \cite{gao99b}.

In a multichannel theory that takes into account the hyperfine structures
of atoms (starting from the formulation in \cite{gao96}), 
the concepts of AQDT, and the concept of $l$-independence in particular, 
remain unchanged and lead to an even greater 
reduction in the number of parameters required for a complete 
characterization of the system \cite{gao03b,gaoup}

We emphasize that AQDT is an exact formulation that does not require 
either the energy-independence or the $l$-independence of $K^c$. 
It is simply the best framework, especially conceptually, to take 
advantage of them when they are there ($\beta_n\gg r_0$). 
The parameterizations 
that we often use to extract universal behaviors should not 
distract from the fact that AQDT also provides an excellent 
platform for exact numerical calculations, whether single 
channel \cite{gao03a} or multichannel. 
This is especially true close to the dissociation limit, where matching
to analytic solutions for $-C_n/r^n$ potential to obtain 
$K^c(\epsilon,l)$ converges much faster than
matching to free-particle solutions to obtain the standard $K$ matrix.
The calculations in \cite{gao03a} are all based on this platform.
[AQDT may even be the best method to calculate the
scattering length when it is close to infinity.]

A major task of AQDT is, of course, finding the reference functions.
This is in general highly nontrivial \cite{gao98a,gao99a}, especially 
analytically. No solution is yet available for $n=5$ at $\epsilon\ne 0$. 
This difficulty is however not a problem here as we need only the
zero-energy reference functions, which can be easily found
for arbitrary $n$ and $l$ \cite{mal61,gao01} 
\begin{eqnarray}
f^c_{\epsilon_s=0 l}(r_s) &=& 
	[2/(n-2)]^{1/2}r_s^{1/2}[J_{\nu_0}(y)\cos(\pi\nu_0/2) \nonumber\\
	& &-Y_{\nu_0}(y)\sin(\pi\nu_0/2)] , 
\label{eq:fc0}\\
g^c_{\epsilon_s=0 l}(r_s) &=& 
	-[2(n-2)]^{1/2}r_s^{1/2}[J_{\nu_0}(y)\sin(\pi\nu_0/2) \nonumber\\
	& &+Y_{\nu_0}(y)\cos(\pi\nu_0/2)] ,
\label{eq:gc0}
\end{eqnarray}
where $\nu_0 = (2l+1)/(n-2)$. With these reference functions,
the zero-energy wave function can be written either as
\begin{equation}
u_{\epsilon=0 l}(r) = A_{l}[f^c_{\epsilon_s=0 l}(r_s) 
	- K^c(0,l) g^c_{\epsilon_s=0 l}(r_s)] \;,
\label{eq:wfnz1}	
\end{equation}
or as
\begin{equation}	
u_{\epsilon=0 l}(r)	= A'_{l} r_s^{1/2}[J_{\nu_0}(y)\cos(\alpha_l) 
	+Y_{\nu_0}(y)\sin(\alpha_l)] \;,
\label{eq:wfnz}
\end{equation}
where $\alpha_l=\pi[\mu^c(0,l)-lb]$ with the quantum defect
$\mu^c(\epsilon,l)$ being defined in terms of $K^c(\epsilon,l)$ 
by Eq.~(\ref{eq:mucdef}). 

The parameters $K^c$ and $\mu^c$ both represent the same physics.
$K^c$ is more convenient in computation, while $\mu^c$ is able to
represent all quantum systems in a finite parameter
space of $[0,1)$. In comparison, $K^c$ can take any value 
from $-\infty$ to $+\infty$.

Equation~(\ref{eq:NlHST}) is simply a result of node-counting
the wave function, given exactly by Eq.~(\ref{eq:wfnz}), from $r_0$ to infinity 
($y=0$ to $y_0$) \cite{wat96}. Equation~(\ref{eq:KcHST}) is obtained simply 
by imposing the boundary condition $u_{\epsilon=0 l}(r=r_0)=0$.

Equation~(\ref{eq:a0sKc}) is obtained by comparing the asymptotic
behavior the $u_{\epsilon=0 l}(r)$ (for $l=0$) at large $r$ with the 
corresponding expansion that defines the $s$ wave scattering length.
\begin{equation}
u_{\epsilon=0 l=0}(r) \rightarrow A(r-a_0) \;.
\end{equation}

The derivation of Eq.~(\ref{eq:Nlu}) in AQDT is straightforward.
In the limit of $r_0\ll \beta_n$, the number of nodes of the
zero-energy wave function inside $r_0$ is an $l$-independent 
constant [to a degree measured by the $l$-independence of 
$K^c(0,l)$]. Counting the number of nodes of the outside
wave function, Eq.~(\ref{eq:wfnz}), from $r_0$ to infinity 
($y=0$ to $y_0$) \cite{wat96}, and ignoring the $l$-dependence of
$\mu^c(0,l)$ leads to Eq.~(\ref{eq:Nlu}).
From this derivation, it is clear that deviation from the universal
behavior is measured by the degree to which $\mu^c(0,l)$ or
$K^c(0,l)$ is independent of $l$. 

Having a bound or quasibound state right at the threshold corresponds
to the boundary condition of $u_{\epsilon=0 l}(r)\rightarrow 0$ (a 
finite constant for $l=0$) in the
limit of $r\rightarrow \infty$. Define
\begin{equation}
x_l(\epsilon) \equiv \tan[\pi\mu^c(\epsilon,l)-\pi lb]\;.
\label{eq:xl}
\end{equation}
From Eq.~(\ref{eq:wfnz}), the condition for a bound-state at the 
threshold is clearly
\begin{equation}
x_l(0) = \tan[\pi\mu^c(0,l)-\pi lb] = 0 \;,
\label{eq:xlb}
\end{equation}
which translates into $\mu^c(0,l)=ib$ for having bound or quasibound 
states of angular momenta $l=i+(n-2)j$ right at threshold,
with corresponding wave functions given by Eq.~(\ref{eq:zebwfn}).
In terms of $K^c$, the same condition takes the form of
\begin{equation}
K^c(0,l) = \tan\left[\frac{1}{n-2}\left(l+\frac{1}{2}\right)\pi\right] \;,
\label{eq:Kcb}
\end{equation}
which is a generalization of the condition in \cite{gao00} to the
case of arbitrary $n$. 
Note that the conditions expressed in the form of
Eqs.~(\ref{eq:xlb}) and (\ref{eq:Kcb}) are exact, with no assumption
on the $l$-independence of either parameter. The universal behavior
corresponds to when the $l$-dependence can be ignored 
($\beta_n\gg r_0$).

The $x_l(\epsilon)$ parameter defined by Eq.~(\ref{eq:xl}) 
has also other applications. For example,
for $\mu^c(0,l)>\approx ib$, $x_l(0)>\approx 0$ is a convenient 
expansion parameter for describing bound  
states of angular momenta $l=i+(n-2)j$ that are close to the threshold.
$x_l(\epsilon)$ is also closely related to the
$K^0_l(\epsilon)$ matrix used in \cite{gao98b}, 
simply by $K^0_l(\epsilon)=-x_l(\epsilon)$. With this relation, all the
results of \cite{gao98b} can be rewritten in terms of either $K^c$ or $\mu^c$.
Making use of the $l$-independence of either parameter in these results
leads, for example, to the relation between $p$ and $s$ wave scattering
lengths. This has been left as an exercise, with the answer to be provided
elsewhere. 

Reference \cite{gao98b} offers an important lesson on the importance of
picking reference functions. The $f^0$ and $g^0$ functions in \cite{gao98a,gao98b},
which define $K^0$, differ from $f^c$ and $g^c$ only by trivial linear
transformation. 
But because the resulting $K^0_l(\epsilon)=-\tan[\pi\mu^c(\epsilon,l)-\pi lb]$
did depend on $l$, relationships between scattering and bound spectra 
of different angular momenta were not recognized until much later 
\cite{gao01,gao00}. Reference \cite{gao98b} was only able to take advantage
of the energy-independence of $K^0_l$ to show, for example, that the 
effective range and the scattering length are not independent, 
but are related in a way determined by the long-range interaction.
The same conclusion was also reached independently by
Flambaum \textit{et al.} using a different approach \cite{fla99}.  

\section{Derivation of the results for LJ(n,2n-2) potentials}
\label{sec:LJnDer}

The analytic results for $N_{\text{LJn}}(l)$, Eq.~(\ref{eq:NlLJn}), 
and $K^c_{\text{LJn}}(0,l)$ given in \cite{gao03a} 
[and therefore $\mu^c_{\text{LJn}}(0,l)$, and of course $a_{0s}$]
are derived from the \textit{zero}-energy
solution of the radial Schr\"{o}dinger equation, Eq.~(\ref{eq:rsch}),
for the class of potentials defined by 
Eq.~(\ref{eq:LJn}). Instead of giving all the boring math details,
we prefer to simply note its relationship to the harmonic
oscillator solution, as they have the same underlying mathematical
structure.

Upon a transformation $x=(r/\beta_n)^\alpha$ and
$u_l(r) = x^{-(\alpha-1)/(2\alpha)}v_l(x)$ with $\alpha=-(n-2)/2$,
the corresponding equation 
at {\it zero\/} energy becomes
\begin{equation}
\left[-\frac{\hbar^2}{2\mu}\frac{d^2}{dx^2} 
	+ \frac{\hbar^2 \gamma(\gamma+1)}{2\mu x^2}
	+ \frac{1}{2}\mu\omega^2 x^2 - E_{e} \right]
	v_l(x) = 0 ,
\label{eq:ho}
\end{equation}
with $\gamma+1/2 = [2/(n-2)](l+1/2)$. Thus
for the class of potentials given by Eq.~(\ref{eq:LJn}),
the solution of the radial Schr\"{o}dinger equation at {\it zero\/} 
energy is equivalent to
the solution of a 3-D isotropic harmonic oscillator with
an effective angular momentum $\gamma$, a effective
frequency determined by $\hbar\omega = (\hbar^2/2\mu)(2/|\alpha|)
(\beta_{2n-2}/\beta_n)^{n-2}(1/\beta_n)^2$,
\textit{at an effective energy (not zero)}
$E_{e}=(\hbar^2/2\mu)(1/\alpha^2)(1/\beta_n)^2$.
From this correspondence, both results are easily deduced.
For example, the number of bound states is simply the number
of harmonic oscillator states below, and including $E_{e}$.

\bibliography{sac}

\end{document}